\documentclass[prb,twocolumn,showpacs,preprintnumbers,amsmath,amssymb]{revtex4}
\usepackage{graphicx}% Include figure files
\usepackage{dcolumn}% Align table columns on decimal point
\usepackage{bm}% bold math
\newcommand \be{\begin{eqnarray}}
\newcommand \ee{\end{eqnarray}}

\begin{document}
\title{Reduction of surface coverage of finite systems 
due to geometrical steps}
        \author{K. Morawetz$^{1,2}$, C. Olbrich$^{1}$, S. Gemming$^{1,3}$ and M. Schreiber$^1$}
\affiliation{$^1$Institute of Physics, Chemnitz University of Technology, 
09107 Chemnitz, Germany}
\affiliation{$^2$Max-Planck-Institute for the Physics of Complex
Systems, N{\"o}thnitzer Str. 38, 01187 Dresden, Germany}
\affiliation{$^3$Forschungszentrum Rossendorf, PF 51 01 19, 01314 Dresden, Germany}

\begin{abstract}
The coverage of vicinal, stepped surfaces with molecules is simulated with 
the help of a two-dimensional Ising model including local distortions and
an Ehrlich-Schwoebel barrier term at the steps. An effective two-spin model is capable 
to describe the main properties of this distorted Ising model. It is employed
to analyze the behavior of the system close to the critical points.
Within a well-defined regime of bonding strengths and Ehrlich-Schwoebel barriers we 
find a reduction of coverage (magnetization) at low temperatures due to the 
presence of the surface step. This results in a second, low-temperature 
transition besides the standard Ising order-disorder transition.
The additional transition is characterized by a divergence of the
susceptibility as a finite-size effect. Due to the
surface step the mean-field specific heat diverges with a power law.
\end{abstract}
\maketitle

\section{Introduction}
The characterization of phase transitions becomes especially demanding in 
situations where the order parameter is not directly accessible by experiment.
One example is the search for a nuclear liquid-gas phase transition.
A considerable discussion can be found in the literature about the 
possibility to observe a negative heat capacity as one signal of such a 
possible liquid-gas phase transition \cite{CKPS04}. Such negative heat capacities appear in finite systems which are adequately described within the microcanonical ensemble. We report 
here an observation that a transition with a divergent heat 
capacity can occur as a consequence of geometrical distortion 
rather than due to a phase transition even in a canonical treatment \cite{FF67}. 
This may shed some light on the nature 
of observed signals. 

One meets a similar situation when describing the coverage of surfaces with molecules. 
There it is interesting to distinguish signals caused by phase transitions between 
different adsorbate arrangements 
from signals due to structural transitions at local deviations from 
the ideal surface geometry.  
Different surface defects have been studied within Ising models \cite{CKPPS00}
by density renormalization methods as well as Monte Carlo techniques. Non-universal 
features were observed and the critical exponent of the magnetization was found 
to be near 1/2 for infinite systems. A review on the vast literature about phase transitions in inhomogeneous systems can be found in \cite{IPT93}. 

We investigate here a finite-size two-dimensional Ising model 
suitable to simulate the coverage of surfaces by molecules. 
%Along one line a surface step is introduced 
%which will be described by a different coupling 
%constant. 
While the explicit simulation with 
realistic parameters was described in [\onlinecite{LZEGSOS06}] we concentrate here
on principal results how the surface modification is 
influencing the transitions and the critical exponents. We suggest 
that the occurrence of divergent (or negative) heat capacities is not a 
unique signal of a phase transition but can occur due to the geometrical 
distortion of the system 
accompanied by anomalous exponents, which even fulfills the scaling hypothesis.   

The two-dimensional Ising model belongs to the most studied models. 
For an overview see [\onlinecite{KKG83}].
The exact solution \cite{O44} shows a phase transition with a critical behavior:

\begin{tabular}{lcll}
spontaneous magnetization & $M$ & $\sim$ & $ |T-T_c|^\beta$ \cr
magnetic field dependence & $H$&$\sim$& $|M_{T=T_c}|^\delta$ \cr
susceptibility & $\chi$&$\sim$&$ |T-T_c|^{-\gamma}$ \cr
specific heat & $c_H$&$\sim$&$ |T-T_c|^{-\alpha}$.
\end{tabular}

\noindent Two exponents are exactly known, i.e. $\beta=1/8$ [\onlinecite{Y52}] and 
$\gamma=1\frac 3 4$ [\onlinecite{A73}]. From asymptotic expansions and strong numerical 
evidence one has furthermore $\alpha=0$ and $\delta=15$ [\onlinecite{GD70}] where the 
specific heat diverges logarithmically. Weiss' mean-field approximation 
instead leads to $\alpha=0$, $\beta=\frac 1 2$, $\gamma=1$, and $\delta=3$ 
[\onlinecite{JM73}]. Both sets of critical exponents fulfill the inequalities
[\onlinecite{R63}] $\alpha+2 \beta+\gamma \ge 2$  and [\onlinecite{G65}] 
$\alpha+\beta(1+\delta)\ge 2$  
known as scaling hypothesis. These scalings are determining 
the corresponding universality classes with specific scaling functions for 
the magnetic field dependence of the magnetization \cite{GD70,K90}.
The universality of this phase transition in two dimensions has been experimentally 
confirmed \cite{BWVRMP95}. Recently, the universality has been investigated with 
respect to finite size scaling \cite{RTMS94} and oscillating fields \cite{KWRN00}. 

Modifications of the scaling relation due to surface defects have been studied extensively, see citations in [\onlinecite{BH74}]. Let us only mention some of the results found in print. The divergences of the specific heat for free and ferromagnetic boundaries in different Ising models have been studied for 40 years \cite{FF67}. The effect of a surface in an Ising model induces spatial correlations which could be treated with the help of a Ginzburg-Landau equation \cite{M71}. The two-spin correlations induced by a line defect in an Ising square lattice are considered with the help of two-particle correlation functions \cite{CP80,KYP85}. Many-point correlation functions along a modified bond have been calculated as well \cite{K81}. The critical exponents for the magnetization of a line defect are known analytically \cite{B79,B82,B00}.

We will present here a quadratic Ising model with a line defect and an additional change of the magnetic field along the line known as Ehrlich-Schwoebel barrier. This can mimic the surface coverage with molecules in the presence of an additional step. First we explain the model and present the numerical results. We will find that the Ehrlich-Schwoebel barrier induces an additional transition. Then in the third chapter we show that the commonly used standard mean-field model fails to explain the observations quantitatively. An effective model is suggested which accounts for the basic results. This effective model is then discussed in chapter IV with respect to the critical exponents and is compared to the mean-field exponents of the standard Ising phase transition.

\section{Ising model with surface step}
In order to simulate the coverage of 
surfaces by molecules we imagine this surface as an $N \times N$ square lattice with a 
straight step across the middle of the lattice.  
The spin-up states describe a molecule sticking to the surface while the spin-down 
states describe the absence of a bound molecule.
The interaction with the $j=4$ neighboring surface molecules is described 
by the coupling constant $J$. 
Across the surface step we choose a different coupling constant 
%$J'$, which is related to $J$ by geometrical scaling as 
$J'=J/\sqrt{2}$. 
The interaction of the surface 
molecules with the substrate background is modeled in analogy to the spin coupling 
with an external field. Therefore we shall use the external magnetic field as a 
synonym for the coupling of molecules with the background. At the sites 
adjacent to the step the magnetic field is augmented by an additional term,
$H_s > 0$. $H_s$ models the Ehrlich-Schwoebel barrier, which impedes the diffusion of 
adsorbates across surface steps. On top of the step edge $H_s$ is added to $H$, hence it
locally reduces the attractive adsorbate-substrate interaction and mimics 
the lower density of coordination sites on top of the step edge. From below,
$H_s$ is subtracted, thus it enhances the adsorbate-substrate interaction and
models the higher number of coordination sites along the step. 
Motivated by the results of Ehrlich and Schwoebel on the stability of step arrays we chose the 
attractive and repulsive parts of the barrier equally high.
Thus, the Ising Hamiltonian for the stepped square lattice reads:
\be
-{\cal H} = 
    \sum_{i,j} J s_i s_j + \sum_{i',j'} J' s_{i'} s_{j'} +
    \sum_{i',j} J s_{i'} s_j  \nonumber\\
    +\mu_0 H \sum_{i} s_i + \mu_0 \sum_{i'} (H \pm H_s) s_{i'} ,
\label{model}
\ee
where $i$ sums over all $N(N-2)$ terrace sites, $i'$ over the step sites,
$j$ over all neighbors with coupling $J$ and $j'$ over all neighbors with coupling $J'$. Without fields $H$ and $H_s$ this Hamiltonian is an Ising square lattice with a ladder defect \cite{IPT93} and the exact critical exponent has been derived \cite{B79} to be
\be
\beta_{\rm H=0}={1\over 2 \pi^2} {\rm arccos}^2 \left (-{\rm tanh} {2(J-J')\over k_B T} \right ).
\label{exap}
\ee
This shows that the critical exponent becomes dependent on the coupling strength and is therefore non-universal.

\begin{figure}[h]
\includegraphics[width=8.2cm]{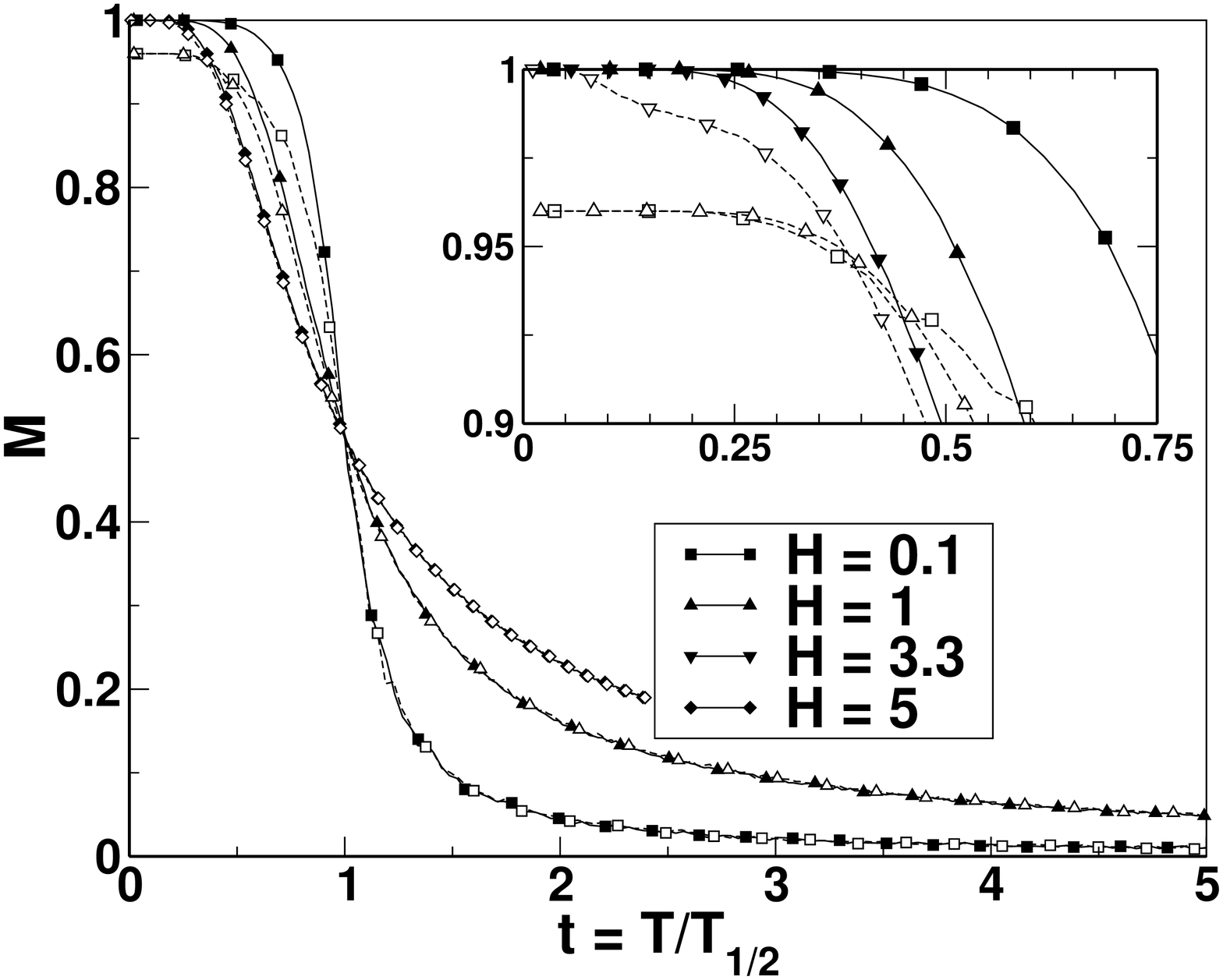}
\caption{Magnetization versus temperature scaled by the half-width temperature where the magnetization is $1/2$ for different external magnetic fields. The Ising lattice (50 $\times$ 50) with a step (open symbols and dashed lines) is compared with the one without a step (closed symbols and solid lines). The parameters are $J=1$, $J'=1/\sqrt{2}$ and $\mu_0H_s=5$. Error bars are of or less than the size of the symbols.}\label{fig1}
\end{figure}

We solve the two-dimensional Ising model with the standard Metropolis 
scheme and the magnetization is now employed as measure for the surface coverage 
with molecules, plotted in Fig.~\ref{fig1}. 
One sees that with increasing external field (or coupling of molecules to the 
background) a smearing of the standard Ising phase transition is obtained,
which results in high temperature tails. This effect is well studied and 
experimentally confirmed \cite{BWVRMP95}. The ferromagnetic transition occurs only for vanishing magnetic fields. 
Fig.~\ref{fig1} also compares the solution of the 
two-dimensional Ising model with and without a surface step: 
For low values of the external magnetic field the step leads to
a characteristic reduction of the magnetization at low temperatures.

\section{Construction of an effective mean-field model}
\subsection{Local mean-field model}
This behavior can be understood by an effective two-spin model. We describe briefly in the following that
the standard mean-field model, as e.g. used in the appendix of [\onlinecite{BH74}], fails. 
For a lattice size of $N\times N$ spins and periodic boundary
conditions, the system is homogeneous in the direction parallel to the step,
thus we can restrict our considerations
to the direction perpendicular to the step. We distinguish $N-2$ normal Ising
spins and $1$ spin at each side of the step, for which the modified coupling 
constant $J'$ and the Ehrlich-Schwoebel barrier term $H_s$ are taken into account \cite{SKR00}. 
All three kinds of spin experience an effective mean-field. 
We will denote the normal mean spin with $s$ and the mean spins at
the step with $s^\pm$ according to the sign of the Ehrlich-Schwoebel 
barrier term, $\pm H_s$. $N-4$ of the sites occupied by normal
spins see a mean-field $\tilde H$ consisting of the external field $H$ and
the interaction with 4 neighbors, $4 J$. The remaining
two of the $N-2$ normal spins interact only with 3 normal spins and with one spin
at the step. Therefore we have
\be
\mu_0 \tilde H=\mu_0 H+{N-4\over N-2} 4 J s+{2 \over N-2} (3 J s + J{s^++s^-\over 2}).
\label{hh}
\ee 
The spins along the step have two interactions with the same kind of spins,
 $2 J s^\pm$, one neighbor with normal
coupling, $J s$, a contribution from the coupling across the step,
 $J' s^\mp$, and an interaction with the substrate of $H \pm H_s$. This
results in
\be
\mu_0 \tilde H^\pm=\mu_0 H\pm \mu_0 H_s+2 J s^\pm+J's^\mp+J s.
\label{hpm}
\ee 

The partition function is then trivially written as
\be
z&=&[ 2 \cosh{(\beta \mu_0 \tilde H)}]^{(N-2) N}
\nonumber\\&&\times
[ 2 \cosh{(\beta \mu_0 \tilde H^+)}]^{N}
[ 2 \cosh{(\beta \mu_0 \tilde H^-)}]^{N}
\label{part}
\ee
with the inverse temperature $\beta=1/k_BT$.
The mean spins are calculated by expressions of the statistical averages,
%\be
$s=
%\!\!{k_B T\over \mu_0 n}{\partial  \ln{z}\over \partial \tilde H}=
\tanh{(\beta \mu_0 \tilde H)}
$
and
%\nonumber\\ \qquad 
$s^\pm
%&=&\!\!{k_B T\over \mu_0 n}{\partial  \! \ln{z}\over \partial  \tilde H'}
=\tanh{(\beta \mu_0 \tilde H^\pm )}.
%\nonumber\\&&
%\label{mean0}
$
%\ee
(\ref{hh}) and (\ref{hpm}) represent 
the self-consistent mean-field equations for $s$ and $s^\pm$.
This mean-field result is exactly equivalent to the Bragg-Williams method by
minimizing the Gibbs functional and assuming that the many-spin correlation
function factorizes into single-spin ones. Such mean-field
equations for open surface defects have been investigated in [\onlinecite{M71,BH74}].

First it is instructive to solve this equation in the limit of zero temperature.
Then one gets the values of the mean spins $\pm 1$ according to the sign
of the mean fields (\ref{hh}) and (\ref{hpm}). Consequently the total mean spin 
\be
M=\left ( 1-\frac 2 N \right ) s+\frac 1 N (s^++s^-)
\label{s0}
\ee
approaches the reduced value 
$M=1-{2\over N}$ for $T\to 0$ if 
%\be 
$\tilde H>0 \,\,  {\rm and} \, \, (\tilde H^+ \gtrless 0\,{\rm and}\, \tilde H^- \lessgtr 0)
$
.
%\label{cond0} 
%\ee
Therefore, as seen in Fig.~\ref{fig1} the reduction is $1-2/50=0.96$ at low
temperatures and low external fields.  Discussing the different cases and
taking into account that the partition function assumes the maximum one deduces
that this reduction happens if and only if
\be
\mu_0 H+3J + J'<\mu_0 H_s
\label{cond1}
\ee
as outlined in the appendix.
Though this mean-field model obviously describes the reduction qualitatively the actual numbers 
do not agree with the simulation result. 
Therefore we can conclude that the standard mean-field model is not able to
describe the effect quantitatively. This is understandable since surface defects induce nonlocal correlations \cite{IPT93}. These nonlocal correlations result in a spatial dependence of the magnetization on the distance from the step on the surface. This can be modeled by a Ginzburg-Landau equation as derived in [\onlinecite{M71}].

\subsection{Effective mean-field model}
A better match with the numerical data is obtained for an effective two-spin
model taking into account these nonlocal correlations in a certain sense. We discriminate now only between normal spins $s$ on attractive sites 
and fictitious spins $s'$ at the repulsive sites with $H-H_s$ along the step. 
In this approach, each row across the terrace contains $N-1$ sites with normal 
spins $s$ and the repulsive site with spin $s'$. An energy-conserving mapping 
of the intuitive three-spin model described above onto this simplified 
two-spin model is possible by setting $s^+ = s$ and $s^- = s'$. 
This mapping relies on the following considerations: 
The presence of one and only one normal spin type $s$ 
is only guaranteed if the effective field is homogeneous on the terrace sites.
On the other hand, the two-spin model 
explicitly accounts only for the repulsive part of the Ehrlich-Schwoebel barrier 
along the step, and omits the attractive part. Yet, the attractive part of the 
Ehrlich-Schwoebel barrier must not be neglected, as it is employed to obtain 
the data from the numerical simulations. 
The only solution consistent with both requirements is to distribute $H_s$ evenly over
the terrace as an overall enhancing field of the strength $H_s/(N-1)$. When summing
over all $N-1$ terrace sites in the Hamiltonian, the same total energy 
of the system results for the two-spin and the three-spin models.

The condition for the second, low-temperature transition can then
be obtained by calculating the site energy $E'$ of the position with spin $s'$
as total energy difference between the three-spin model and the attractive part
of the two-spin model. This procedure yields for the site energy
\be
E' = \mu_0H(s^++s^--s) + \mu_0H_s(s^+-s^--s)  \nonumber\\
   + 2J\left [(s^+)^2+(s^-)^2+ss^++ss^--2s^2+\frac{1}{2}ss'\right ]  \nonumber\\
   + 2J'(s^+s^--\frac{1}{2}ss').
\ee

Equating $s = s^+ = 1$ and $s' = s^- = -1$ one obtains for the site energy
$E' = -\mu_0H + \mu_0H_s - J - J'$ for the effective spin $s'$ in anti-parallel
orientation with respect to $s$. This orientation is favorable, if $E' < 0$,
hence the condition for a reduction of the magnetization reads:
\be
\mu_0 H+J + J'<\mu_0 H_s
%\qquad T_c'={J'\over 2 J} T_c
.
\label{cond}
\ee
Indeed, numerical simulations for different parameter sets  $J,H,H_s$ 
confirm this result. Hence, the effective two-spin model is employed for the further analysis of the
numerical simulations.

The mean field of the normal spins is calculated analogously to (\ref{hh})  
\be
\mu_0 \tilde H=\mu_0 H+{N-3\over N-1} 4 J s+{6 s+2 s' \over N-1} J  \equiv \mu_0 H+j J s.
\label{hh1}
\ee 
The fictitious spins differ from the normal ones by a constant $s' = c +s$ where $-2<c<2$. For a given value of $T$, each spin $s$ has therefore a mean field $j J s  = 4 J s+  2 J  c/(N-1)\approx 4 J s $ with a maximal relative error of $1/(N-1)$ which we can neglect in the following.

The effective spins along the step are described by
a mean field consisting of the linear combination of the couplings $J,J'$ with
the spins $s,s'$. Taking into account (\ref{cond}) and that a possible second transition
can only occur at a temperature $T_c'\sim J'$ as well as that for $J=J'$ and
$H_s=0$ the normal Ising model should reappear, we obtain uniquely the mean field of the effective spins as
\be
\mu_0 \tilde H' = \mu_0 H-\mu_0 H_s+(J+J') s+ 2 J' s',
\label{H'}
\ee 
the derivation of which is outlined in the appendix.
The partition function can again be trivially written and
the mean spins are
\be
s'&=&\tanh{[\beta ( \mu_0 (H-H_s)+(J+J') s+2 J' s')]}\nonumber\\
s&=&\tanh{[\beta (\mu_0 H+ j J s)]},
\label{mean}
\ee
representing the self-consistent mean-field equations for $s$ and $s'$.
We find from (\ref{mean}) that the total mean spin 
\be
M=\left ( 1-\frac 1 N \right ) s+\frac 1 N s'
\label{s}
\ee
approaches the reduced value $M=1-2/N$ in the limit of zero temperature
if and only if the condition (\ref{cond}) is fulfilled as shown in the 
appendix.

\begin{figure}[h]
\includegraphics[width=9cm]{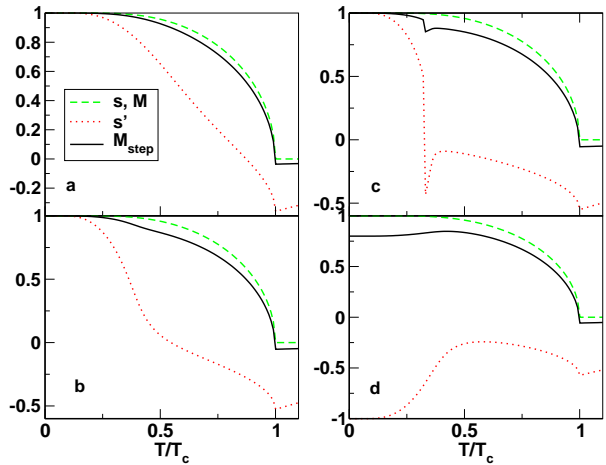}
\caption{The mean spins $s$ (dashed line) and $s'$ (dotted line) as solution of (\protect\ref{mean}) versus temperature together with the total mean magnetization, M$_{\rm step}$, (solid line) for $\mu_0 H_s=0.10,0.16,0.17,0.18$ (a-d). The magnetization without step agrees with $s$. The lattice size is $N=10$, the couplings $J=0.1$ and $J'=J/\sqrt{2}$, and the external field $H=0$.}\label{fig2} 
\end{figure}

The solution of Eq. (\ref{mean}) versus temperature is plotted in
Fig.~\ref{fig2} for different values of $\tilde H'$. We see the characteristic
reduction of the effective magnetization to $1-2/10=0.8$. This reduction
occurs as long as $0.1+0.071=0.171<\mu_0 H_s$ according to (\ref{cond}). 
Fig.~\ref{fig2}(c) displays that slightly below the critical value 
($\mu_0 H_s=0.17$) we do not have a reduction at $T=0$ but a
sharp drop of the magnetization around $T/T_c=2 J'/ jJ=0.35$. This is related
to a maximum in the specific heat at a second critical temperature
$T_c' \approx 0.35 T_c$ besides the usual Ising transition temperature $T_c$
as shown in Fig.~\ref{heat}. The same second transition appears in
Fig.~\ref{fig1} where $\mu_0H_s=5$ and consequently the reduction occurs
as long as $\mu_0H<3.29$.

\section{Mean-field critical exponents of new transition}
We can understand the second transition by expanding (\ref{mean}) for
 low fields $H$ and calculating the susceptibility 
\be
\chi={\partial M \over \partial H}|_{H=0}.
\label{chi}
\ee
From (\ref{mean}) we obtain
\be
{\partial s'\over \partial H}\!&=&\!\beta(1\!-\!s'^2) \left ( \mu_0\!+\! ({T_c \over j}+{T_c'\over 2}) {\partial s\over \partial H}\!+\!T_c'{\partial s'\over \partial H} \right )
\nonumber\\
{\partial s\over \partial H}&=&\beta(1-s^2) \left ( \mu_0+T_c {\partial s\over \partial H}\right )
%\nonumber\\
\label{deriv}
\ee
which is easily solved and employed to calculate (\ref{chi}). 
We discuss this susceptibility explicitly near the two transitions.
At the usual phase transition $T_c=j J/k_B$ where $s=0$ 
we have
\be
\chi|_{s=0}\!\!&=&\!\!\left ( \!{N\!-\!1 \over N}\!-\! {1\!-\!s'^2\over N}{T\!+\!\frac 1 2 T_c'\!+\!({1\over j}\!-\!1)T_c\over T\!-\!(1\!-\!s'^2) T_c'}\right ){\mu_0/k_B\over T\!-\!T_c}\nonumber\\
\lim\limits_{N\to\infty} \chi|_{s=0}\!\!&=&\!{\mu_0/k_B\over T-T_c} 
\label{sus}
\ee
and the typical critical exponent $\gamma=1$ occurs for finite and infinite 
lattices. The step of the surface does not change the critical scaling of the susceptibility.

Near the second transition at $T_c'=2 J'/k_B$  where $s'=0$ we obtain 
\be
\chi|_{s'=0}&=&{\mu_0\over k_B}\left ( {1-N \over N} {1-s^2\over T-(1-s^2)T_c}\right .
\nonumber\\&&+\left . {1\over N}{T+ (1-s^2)((1/j-1)T_c+\frac 1 2 T_c')\over (T-(1-s^2) T_c)(T-T_c')}\right )\nonumber\\
\lim\limits_{N\to\infty} \chi|_{s'=0}&=&-{\mu_0(1-s^2)/k_B\over T-(1-s^2)T_c}
\label{sus1}
\ee
and for a finite lattice ($N<\infty$) we see that $\chi\sim 1/(T-T_c')$.
Consequently, at the second critical temperature, $T_c'$, the
susceptibility diverges and a sharp drop of magnetization occurs with the
critical exponent $\gamma'=1$. This second critical temperature
does not appear for infinite lattices since the term with $1/(T-T_c')$
vanishes in the limit $N\to\infty$.
We hence conclude that the second transition occurs due to 
the finite spacing between two adjacent surface steps.

Even the quantitative behavior of the mean-field
model agrees remarkably well with the numerical solution if we scale to the corresponding
half-width temperatures as can be seen in Fig.~\ref{vergleich}.
Especially the low-temperature behavior and the drop at the second,
low-temperature transition at $T_c'$ are well described. 
Since the mean-field approximation does not yield the correct critical exponents
of the standard order-disorder transition of the planar two-dimensional Ising model 
it is in accordance with previous findings that deviations occur for temperatures 
higher than $T_{1/2}$.

\begin{figure}[]
\includegraphics[width=8.2cm]{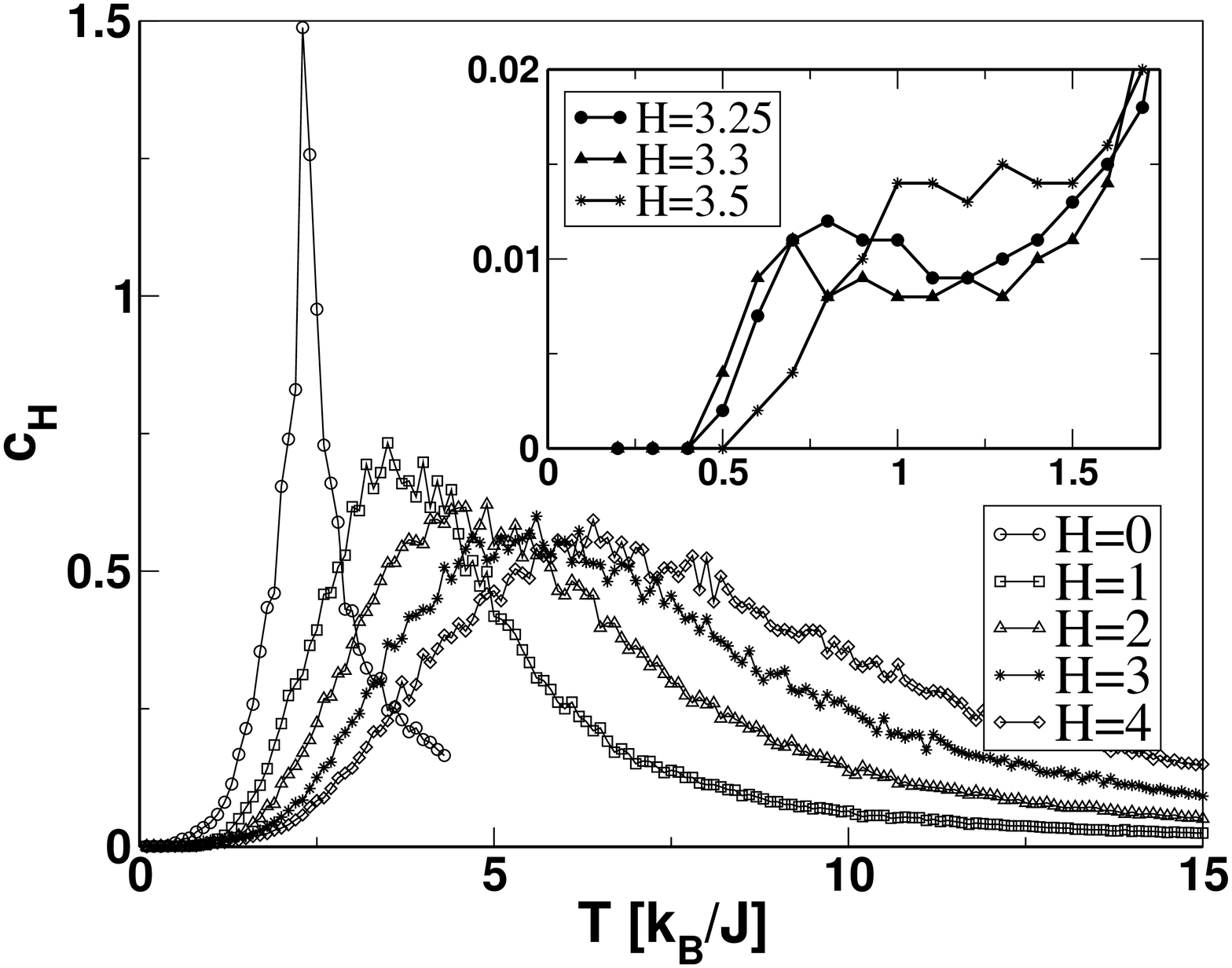}
\caption{The specific heat for parameters of Fig. \protect\ref{fig1} 
for the numerical data of the Ising model with a step.}\label{heat} 
\end{figure}

In order to substantiate the picture of a second, low-temperature transition 
we investigate the remaining critical exponents. We find the magnetic 
field dependence of the magnetization by rewriting (\ref{mean})
\be
{\mu_0 H\over k_B T}&=&-{T_c\over T}s+\frac 1 2 \ln\left ({1+s\over 1-s}\right )
\label{sH}
\ee
and near $T_c$
\be
{\mu_0 H\over k_B T_c}&=& \frac 1 3\left ({N\over N-1} M -{s'\over N-1}\right )^3+o (s^5)
\ee
where (\ref{s}) has been used. We see that $\delta=0$ for finite lattices while
for $N\to\infty$ we obtain the established value $\delta=3$ of the standard Ising model.
In the same way we obtain near $T_c'$
\be
{\mu_0(H\!-\! H_s)\over k_B T_c'}\!&=&\!-\!\left ({T_c\over j T_c'} +\frac 1 2\right) s\!-\!s'\!+\!\frac{T}{2T_c'} \ln \! \left ({1\!+\!s'\over 1\!-\!s'}\right )\nonumber\\
&=&\!-\left ({T_c\over j T_c'} +\frac 1 2\right)
 {N\over N\!-\!1} M+o(s')
\ee
and we see that in both, finite and infinite lattices (with $H_s=0$)
we have $\delta'=1$, which is different from the standard Ising model.
The finite Ehrlich-Schwoebel barrier, $H_s\ne 0$, leads to $\delta'=0$.

We find the spontaneous magnetization for $H=0$ near $T_c$ from (\ref{sH})
\be
{T_c/T}=1+{s^2\over 3}+o(s^4)
\label{stc}
\ee 
which results with (\ref{mean}) in
\be
M={s'\over N} +{N-1\over N}\sqrt{3 \left ( {T_c\over T}-1\right )}
\label{Ms}
\ee
such that we have $\beta=\frac 1 2$ for the infinite-size limit and
 $\beta=0$ for the finite-size case.
Near $T_c'$ we obtain analogously $
{-\mu_0 H_s}+(\frac 1 j T_c+\frac 1 2 T_c')s +{T_c'} s'=s' T+o(s'^3)
%\label{sptc}
$ which leads to
\be
\!M\!=\!\frac{N\!-\!1}{N} {T\!-\!T_c'\over \frac 1 j T_c\!+\!\frac 1 2 T_c'}s'\!+\!\frac 1 N s'\!+\!{N\!-\!1\over N} {\mu_0 H_s \over k_B(\frac 1 j T_c\!+\!\frac 1 2 T_c')}
\ee
and $\beta'=0$ for finite-size lattices independent of the Ehrlich-Schwoebel barrier. The same exponent appears for infinite size since according to (\ref{s}) the anomalous spins $s'$ do not contribute to the magnetization in the infinite limit.

\begin{figure}[h]
\includegraphics[width=8.2cm]{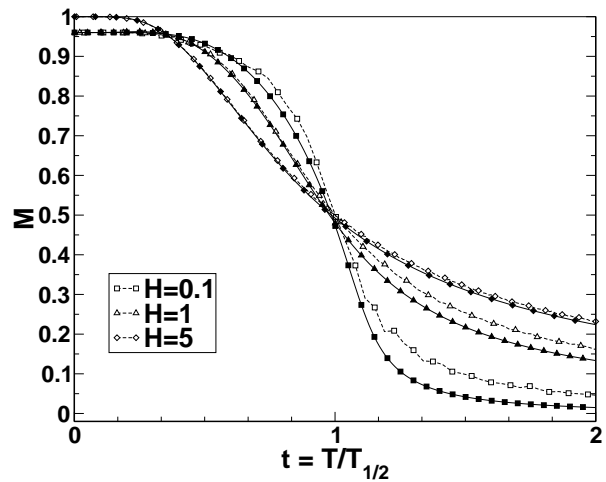}
\caption{The magnetization of the Ising model (open symbols and dashed line) with a step as in Fig.~\protect\ref{fig1}, together with the solution of (\protect\ref{mean}) (closed symbols and solid lines).}\label{vergleich} 
\end{figure}

Besides the divergence at $T_c$ the specific heat shows a second maximum 
at $T_c'=2 J'/J T_c$ for fields fulfilling (\ref{cond}) as can be seen in
Fig.~\ref{heat}. The interesting leading order near the critical points
at $H=0$, where we have $s=\sqrt{3(T_c/T-1)}$ from (\ref{stc}), reads
\be
&&c_H/k_B={\sqrt{3} (N-1) s' T_c^{3/2}(\frac 1 j T_c+\frac 1 2 T_c') \over 4 N^2 (T_c -(1-s'^2) T_c')} {1\over (T_c-T)^{3/2}}
\nonumber\\&&
+{o(\frac 1 N)\over T\!-\!T_c}\!+\!{o(\frac 1 N)\over \sqrt{T\!-\!T_c}}\!-\!\frac 9 8 \!+\!o(\frac 1 N)\!+\!o(\sqrt{T-T_c})
%\nonumber\\&&
\ee
which leads to $\alpha=3/2$ for the finite and $\alpha=0$ for the infinite case.
Near the other critical point $T_c'$ with $s'\to 0$ we have the leading order
\be
{c_H'\over k_B}&=&{({\mu_0\over k_B} H_s)^2 (1\!-\!N) [a\!+\!T_c' (2 T_c\!+\!j T_c')^2]^2\over a^2 N^2 (T \!-\!T_c')^2}
\!+\!{o(\frac 1 N) \over T\!-\!T_c'}
\nonumber\\&+&
({2 \mu_0\over k_B}H_s jT_c)^2[({2 \mu_0\over k_B}H_s j)^2-(2 T_c\!+\!j T_c')^2] \nonumber\\
&\times& {a\!\!+\!\!2 (T_c\!\!-\!\!T_c')(2 T_c\!\!+\!\!j T_c')^2\over a^3}
\!+\!o(\frac 1 N)\!+\!o(T\!-\!T_c')
\ee
with $a=-4 \mu_0^2 H_s^2 j^2 T_c/k_B^2+(T_c-T_c')(2 T_c+j T_c')^2$. It shows
$\alpha'=2$ for the finite case and $\alpha'=0$ for the infinite case.
In the case with no Ehrlich-Schwoebel barrier ($H_s=0$) the specific heat becomes
$c_H'\sim o(s'^2)$
which shows $\alpha'=0$, and no second transition $s'=0$ occurs for finite or infinite systems.

\begin{table}
\begin{tabular}{l||l|l|l|l||l|l}
& $\alpha$ & $\beta$ & $\gamma$ & $\delta$& $\alpha\!\!+\!\!2 \beta\!\!+\!\!\gamma$ & $\alpha\!\!+\!\!\beta(1\!\!+\!\!\delta)$ \cr\hline &&&&&&\cr
2D Ising (exact)&0 &1/8&7/4&15&2&2\cr
2D Ising (Weiss) &0 &1/2 & 1 & 3&2&2 \cr&&&&&&\cr\hline &&&&&&\cr
$T_c$ $N\!\ne\! \infty$ & 3/2 & 0 & 1 & 0 & 2.5 & 1.5 \cr
$T_c$ $N\!=\!\infty$ & 0 & 1/2 & 1 & 3 & 2 & 2 \cr&&&&&&\cr\hline&&&&&&\cr 
$T_c'$ $N\!\ne\! \infty$ ($H_s\!=\!0$) & 2 (0) & 0 (0) & 1 (1) & 0 (1)& 3 (1)& 2 (2) \cr
$T_c'$ $N\!=\!\infty$ ($H_s\!=\!0$) & 0 (0) & 0 (0) & 0 (0) & 0 (1) & 0 (0) & 0 (0) \cr
 \end{tabular}
\caption{Critical exponents for the two transitions in mean-field approximation. The value in brackets give the results without Ehrlich-Schwoebel barrier.}\label{tab1}
\end{table}

One should note that the divergence of the specific heat appears here in the mean-field model though the numerical data show a mere maximum. This rounding of the divergence is due to the finite size of the lattice and well discussed, see [\onlinecite{FF67}].

The results for the mean-field model are summarized in table~\ref{tab1}. 
We see that for infinite
lattices the presence of the step does not change the exponents of the Ising model
near the normal transition $T_c$.  The finite-size effects lead to a
deviation of all exponents from the result without step except the exponent of
the susceptibility which remains unchanged. Especially the specific heat
changes from logarithmic divergence to power-law divergence. For the
reported second transition the scaling inequalities are fulfilled.
In the infinite-size limit no second transition occurs.

Please note that we compare here the mean-field critical exponents for the new transition arising due to the Ehrlich-Schwoebel barrier. The exact one for the ladder defects without Ehrlich-Schwoebel barrier is well known \cite{B79}, as presented above (\ref{exap}). 

\section{Summary}
For Ising systems on a square lattice with a spatial 
                     distortion we report here that a second, low-temperature 
                     transition occurs besides the standard Ising phase transition.
An analytical two-spin model is capable to describe the main features of such a
distorted finite spin system. The divergent heat capacity appears here due to
the spatial distortion and not due to an Ising phase transition. Therefore,
experimentally recorded signals with divergent heat capacities may not exclusively be interpreted in terms of phase transitions in finite systems. 
When simulating the surface coverage
with molecules the present model is able to describe the main
equilibrium features \cite{LZEGSOS06}, thus it promises an application potential
to the fabrication of nanowires which are created near a surface step.

\appendix
\section{Critical Ehrlich-Schwoebel barrier}

Here we outline the discussion of the critical Ehrlich-Schwoebel barrier where the second transition occurs in a two-spin model if condition (\ref{cond}) is fulfilled. A completely analogous discussion leads to the result for the three-spin model (\ref{cond1}). 

The fictitious spin $s'$ obeys the equation
\be
s'={\rm tanh} \beta (H-H_s+a s+b s')
\ee
and the second transition occurs if $s=1$ and $s'=-1$ since only in this case the magnetization (\ref{s}) is reduced. For zero temperature the ${\rm tanh}$ function shows that $s'=\pm1$ if and only if 
\be
c_\pm=H+a \pm b\gtrless H_s.
\label{app1}
\ee
Since $c_-<c_+$ we have the situation that for $H_s<c_-$ we have $s'=1$ and for $H_s>c_+$ we have $s'=-1$ while for $c_-<H_s<c_+$ both solutions $s'=\pm 1$ exists. In this range the system will take the solution where the partition function becomes maximal. Since the partition function is proportional to ${\rm cosh}(c_\pm-H_s)$ we have $s'=\pm 1$ if 
\be
|c_+-H_s|\gtrless |c_--H_s|.
\ee
Since we considered the range $c_-<H_s<c_+$ we obtain with (\ref{app1})
\be
H+a\gtrless H_s
\ee
as a unique condition where $s=1$ and $s'=\pm 1$ and where the second transition occurs.
\bibliography{spin1}

\end{document}